\documentclass[final,twocolumn,5p,times,10pt,authoryear]{elsarticle}
\usepackage{lineno}
\modulolinenumbers[5]

\journal{Automatica}

\usepackage[american]{babel}
\usepackage[utf8]{inputenc}
\usepackage[T1]{fontenc}
\usepackage{graphicx}
\usepackage{amsmath}
\usepackage{amssymb}
\usepackage{amsthm}
\usepackage{amsfonts}
\usepackage{bm}
\usepackage{hhline}
\usepackage{csquotes}
\usepackage{color}
\usepackage{xcolor}
\usepackage{import}
\usepackage{algorithm}
\usepackage{algpseudocode}
\usepackage{dsfont}
\usepackage{multirow}
\usepackage{soul}
\setstcolor{magenta}

\makeatletter
\usepackage[colorlinks=true,linkcolor=red]{hyperref}
\makeatother

\newtheorem{theorem}{\bf{Theorem}}
\newtheorem{proposition}{\bf{Proposition}}
\newtheorem{definition}{\bf{Definition}}
\newtheorem{remark}{\bf{Remark}}
\newtheorem{corollary}{\bf Corollary}

\allowdisplaybreaks 

\newcommand{\mbf}[1]{\ensuremath{{\mathbf{#1}}}}

\newcommand{\half}{\ensuremath{\frac{1}{2}}}

\newcommand{\eye}[1]{\ensuremath{\mbf{I}_{#1}}}
\newcommand{\eyenoarg}{\ensuremath{\mbf{I}}}
\newcommand{\zeros}[2]{\ensuremath{\bm{0}_{#1\times#2}}}

\newcommand{\real}[1]{\ensuremath{\text{Re}(#1)}}

\newcommand{\realset}{\ensuremath{\mathbb{R}}}
\newcommand{\realsetmat}[2]{\ensuremath{\mathbb{R}^{#1\times#2}}}
\newcommand{\intvalset}{\ensuremath{\mathbb{I}\mathbb{R}}}
\newcommand{\intvalsetmat}[2]{\ensuremath{\mathbb{I}\mathbb{R}^{#1\times#2}}}

\newcommand{\lbound}{\ensuremath{\text{L}}}
\newcommand{\ubound}{\ensuremath{\text{U}}}
\newcommand{\midpoint}[1]{\ensuremath{\text{mid}({#1})}}

\newcommand{\rad}[1]{\ensuremath{\text{rad}({#1})}}

\newcommand{\gzinclusion}{\ensuremath{\triangleleft}}

\newcommand{\ninf}[1]{\ensuremath{\|{#1}\|_\infty}}

\definecolor{orange}{RGB}{255,69,0}

\newcommand{\bibfolder}{Bibliography}
\biboptions{authoryear}

\begin{document}

\begin{frontmatter}

\title{Joint state and parameter estimation based on constrained zonotopes\tnoteref{mytitlenote}}

\tnotetext[mytitlenote]{{\copyright} 2022. This manuscript version is made available under the CC-BY-NC-ND 4.0 license \url{https://creativecommons.org/licenses/by-nc-nd/4.0/}. Work supported by the Brazilian agencies
CNPq (INCT grant no. 465755/2014-3, PQ grant no. 315695/2020-0), CAPES (Grants no. 001 and 88887.136349/2017-00), FAPEMIG (Grant no. APQ-03090-17), and FAPESP (INCT grant no. 2014/50851-0), and the Italian Ministry for Research in the framework of the 2017 Program for Research Projects of National Interest (PRIN) (Grant no. 2017YKXYXJ).}

\author[myfirstaddress]{Brenner S. Rego\corref{mycorrespondingauthor}}
\cortext[mycorrespondingauthor]{Corresponding author, \textit{brennersr7@ufmg.br} }

\author[mysecondaddress]{Diego Locatelli}

\author[mysecondaddress]{Davide M. Raimondo}

\author[myfirstaddress,mythirdaddress]{Guilherme V. Raffo}

\address[myfirstaddress]{Graduate Program in Electrical Engineering, Federal University of Minas Gerais, Belo Horizonte, MG 31270-901, Brazil}
\address[mysecondaddress]{Department of Electrical, Computer and Biomedical Engineering, University of Pavia, Italy}
\address[mythirdaddress]{Department of Electronics Engineering, Federal University of Minas Gerais, Belo Horizonte, MG 31270-901, Brazil}

\begin{abstract}
This note presents a new method for set-based joint state and parameter estimation of discrete-time systems using constrained zonotopes. This is done by extending previous set-based state estimation methods to include parameter identification in a unified framework. Unlike in interval-based methods, the existing dependencies between states and model parameters are maintained from one time step to the next, thus providing a more accurate estimation scheme. In addition, the enclosure of states and parameters is refined using measurements through generalized intersections, which are properly captured by constrained zonotopes. The advantages of the new approach are highlighted in two numerical examples.
\end{abstract}


\end{frontmatter}

\section{Introduction}

Without assuming knowledge of the stochastic properties of unknown variables, 
set-based state estimation methods are able to provide guaranteed enclosures of the system trajectories
in applications affected by bounded uncertainties \citep{Chisci1996,Scott2016}. 
Set-based methods have also been widely used in the parameter identification field as an alternative to stochastic methods, 
since they are able to provide guaranteed enclosures of the model parameters when the uncertain model parameters have unknown stochastic properties. 
Zonotopes have been used to approximate the parametric set 
for discrete-time systems with additive uncertainties in \cite{Bravo2006}, which was later extended to allow multiplicative uncertainties in \cite{Wang2017}. However, both methods are applied only to systems described by regression models, and rely on conservative intersections with strips to refine the parametric set. Intervals have been used in the context of optimal design of experiments in \cite{Vidal2019}, 
to minimize the conservatism of the parametric enclosure. Moreover, a bisection-based interval algorithm has been used in \cite{Rumschinski2010} to deal with non-convex parameter sets using collections of intervals. Nevertheless, intervals are not able to capture dependencies between variables, which may result in conservative enclosures due to  wrapping effect. 

In the literature, parameter identification is typically addressed as a separated problem from state estimation, in which a model is identified off-line. 
Few state estimation strategies in the literature refine online the model parametric uncertainties in order to improve the accuracy of state estimation. Such methodology is referred to as joint state and parameter estimation, 
which enables the simultaneous estimation of both states and model parameters. It allows for a more efficient update of these variables using available measurement, besides taking into account state-parameter dependencies, rather than dealing with two separated problems.  
A Kalman filtering (KF) strategy, based on multi-innovation recursive extended least squares algorithm, has been proposed in \cite{Cui2020} to enhance parameter estimation. However, bias issues introduced by KF make such approaches unreliable in case the assumptions on the stochastic properties of the uncertainties are violated.
Deterministic approaches include Luenberger-based observers \citep{Zhang2020} and set-based interval estimation \citep{raissi2004set}. The latter propose a prediction-update state and parameter estimator suitable for nonlinear continuous-time systems. However, besides not being able to capture the dependencies between states and parameters, the method can lead to high computational complexity due to the use of multiple sets.

The work presented in this note proposes a method for set-based joint state and parameter estimation of discrete-time systems. The strategy extends the algorithms based on constrained zonotopes (CZs) proposed in \cite{Scott2016} and \cite{Rego2021}, to include parameter estimation in a unified framework for the first time. In contrast to interval-based methods\footnote{For comparison purposes, we extend the methods in \cite{Jaulin2001b} and \cite{Alamo2005a} to the proposed framework to include parameter estimation.}, this framework implemented using CZs allows the estimated enclosures to propagate existing dependencies between states and model parameters. Besides, both the state and parameter enclosures (which are unified in our method) are refined using generalized intersections, unlike in zonotope-based estimation methods. These advantages result in a significant improvement in the accuracy of both state and parameter estimation.

\section{Preliminaries} \label{sec:preliminaries}

Consider $Z, W \subset \realset^{n}$, $Y \subset \realset^{m}$, and a real matrix $\mbf{R} \in \realset^{m \times n}$. Let $Z \times W$ be the Cartesian product, and define the linear mapping, Minkowski sum, and generalized intersection, as
\begin{align}
\mbf{R}Z & \triangleq \{ \mbf{R} \mbf{z} : \mbf{z} \in Z\}, \label{eq:pre_limage}\\
Z \oplus W & \triangleq \{ \mbf{z} + \mbf{w} : \mbf{z} \in Z,\, \mbf{w} \in W\}, \label{eq:pre_msum}\\
Z \cap_{\mbf{R}} Y & \triangleq \{ \mbf{z} \in Z : \mbf{R} \mbf{z} \in Y\}, \label{eq:pre_intersection}
\end{align}
respectively. In this note, functions with set-valued arguments will be used to denote the exact image of the set under the function, i.e. $\bm{\mu}(X,W)\triangleq\{\bm{\mu}(\mathbf{x},\mathbf{w}): \mathbf{x}\in X, \ \mathbf{w}\in W\}$. 
In addition, let $\bm{\kappa}$ be a function of class $\mathcal{C}^1$ (i.e., continuously differentiable) and $\mbf{z}$ denote its argument. Then, $\kappa_q$ denotes the $q$-th component of $\bm{\kappa}$, and $\nabla^T_z \bm{\kappa}$ denotes the Jacobian of $\bm{\kappa}$ with respect to \mbf{z}. \emph{Constrained zonotopes} are an extension of zonotopes, defined as in \cite{Scott2016}, capable of describing also asymmetric convex polytopes, while maintaining many of the well-known computational benefits of zonotopes \citep{Kuhn1998}.

\begin{definition} \rm \label{def:pre_czonotopes}
	A set $Z \subset \realset^n$ is a \emph{constrained zonotope} if there exists $(\mbf{G}_z,\mbf{c}_z,\mbf{A}_z,\mbf{b}_z) \in \realsetmat{n}{n_g} \times \realset^n \times \realsetmat{n_c}{n_g} \times \realset^{n_c}$ such that
	\begin{equation} \label{eq:pre_cgrep}
	Z = \left\{ \mbf{c}_z + \mbf{G}_z \bm{\xi} : \| \bm{\xi} \|_\infty \leq 1, \mbf{A}_z \bm{\xi} = \mbf{b}_z \right\}.
	\end{equation}	
\end{definition}

We refer to \eqref{eq:pre_cgrep} as the \emph{constrained generator representation} (CG-rep). Each column of $\mbf{G}_z$ is a \emph{generator}, $\mbf{c}_z$ is the \emph{center}, $\mbf{A}_z \bm{\xi} = \mbf{b}_z$ are the \emph{constraints}, and $\bm{\xi}$ are the \emph{generator variables}. By defining the constrained unitary hypercube\footnote{We use the notation $B_\infty^{n_g}$ for the $n_g$-dimensional unitary hypercube (i.e., without equality constraints). We drop the superscript $n_g$ for $B_\infty(\mbf{A}_z,\mbf{b}_z)$ since this dimension can be inferred from the number of columns of $\mbf{A}_z$.} $B_\infty(\mbf{A}_z,\mbf{b}_z) \triangleq \{\bm{\xi} \in \realset^{n_g} : \ninf{\bm{\xi}} \leq 1,\,  \mbf{A}_z \bm{\xi} = \mbf{b}_z \}$, a CZ $Z$ can be written as 
$Z = \mbf{c} \oplus \mbf{G}_z B_\infty(\mbf{A}_z,\mbf{b}_z)$. We use the compact notation $Z = \{\mbf{G}_z, \mbf{c}_z,\mbf{A}_z,\mbf{b}_z \}$ for CZs, and $Z = \{\mbf{G}_z, \mbf{c}_z\}$ for zonotopes. %
The set operations \eqref{eq:pre_limage}--\eqref{eq:pre_intersection} can be computed exactly with CZs. Let $Z = \{\mbf{G}_z, \mbf{c}_z, \mbf{A}_z, \mbf{b}_z\} \subset \realset^n$, $W = \{\mbf{G}_w, \mbf{c}_w, \mbf{A}_w, \mbf{b}_w\} \subset \realset^n$, $Y = \{\mbf{G}_y, \mbf{c}_y, \mbf{A}_y, \mbf{b}_y\} \subset \realset^m$, and $\mbf{R} \in \realsetmat{m}{n}$. Then, \eqref{eq:pre_limage}--\eqref{eq:pre_intersection} are computed trivially in CG-rep as \citep{Scott2016}
\begin{align}
\mbf{R}Z & = \left\{ \mbf{R} \mbf{G}_z, \mbf{R} \mbf{c}_z, \mbf{A}_z, \mbf{b}_z \right\}, \label{eq:pre_czlimage}\\
Z \oplus W & = \left\{ [ \mbf{G}_z \,\; \mbf{G}_w ], \mbf{c}_z + \mbf{c}_w, \begin{bmatrix} \mbf{A}_z & \bm{0} \\ \bm{0} & \mbf{A}_w \end{bmatrix}, \begin{bmatrix} \mbf{b}_z \\ \mbf{b}_w \end{bmatrix} \right\}, \label{eq:pre_czmsum}\\
Z \cap_{\mbf{R}} Y & = \left\{ [\mbf{G}_z \,\; \bm{0}], \mbf{c}_z, \begin{bmatrix} \mbf{A}_z & \bm{0} \\ \bm{0} & \mbf{A}_y \\ \mbf{R} \mbf{G}_z & -\mbf{G}_y \end{bmatrix}, \begin{bmatrix} \mbf{b}_z \\ \mbf{b}_y \\ \mbf{c}_y - \mbf{R} \mbf{c}_z \end{bmatrix} \right\}. \label{eq:pre_czintersection}
\end{align}

Operations \eqref{eq:pre_czlimage}--\eqref{eq:pre_czintersection} cause a linear increase in the complexity of the CG-rep. Moreover, unlike other set representations (such as ellipsoids, intervals, and zonotopes), all operations \eqref{eq:pre_czlimage}--\eqref{eq:pre_czintersection} are exact using CZs, and therefore, can be computed efficiently and accurately. Efficient methods for complexity reduction of CZs (to enclose a CZ with another one with a fewer number of generators and constraints) are available \citep{Scott2016}. In this note, $\intvalset$ denotes the set of real compact intervals. Let $X \triangleq \{ a \in \realset : x^\lbound \leq a \leq x^\ubound \} \in \intvalset$ be an interval. Then, $\midpoint{X} \triangleq \half(x^\ubound + x^\lbound)$ and $\rad{X} \triangleq \half(x^\ubound - x^\lbound)$. Let $\mbf{N} \triangleq \{N_{ij} \in \intvalset, i \in \{1,\ldots,n\}, j \in \{1,\ldots,m\}\} \in \intvalsetmat{n}{m}$ be an interval matrix. Then, $\midpoint{\mbf{N}}$ and $\rad{\mbf{N}}$ are defined component-wise. In addition, for any bounded $Z$, $\square Z$ denotes the interval hull of $Z$. If $Z$ is a CZ, this operation is performed by solving $2n$ linear programs (LPs) \citep{Scott2016}. In the following, Theorem \ref{thm:ndyn_czinclusion} defines the operation $\gzinclusion(\mbf{J},X)$ for enclosing the product of an interval matrix $\mbf{J}$ with a CZ $X$. When not required, the subscripts of the variables in \eqref{eq:pre_cgrep} will be omitted. 

\begin{theorem} \rm \label{thm:ndyn_czinclusion} \citep{Rego2020}	
	Let $X = \{\mbf{G},\mbf{c},\mbf{A},\mbf{b}\} \subset \realset^m$ be a CZ with $n_g$ generators and $n_c$ constraints, let $\mbf{J} \in \intvalsetmat{n}{m}$ be an interval matrix, and consider the set $S = \mbf{J} X \triangleq \{\hat{\mbf{J}} \mbf{x} : \hat{\mbf{J}} \in \mbf{J}, \mbf{x} \in X\} \subset \realset^n$. Let $\bar{\mbf{G}} \in \realsetmat{n}{\bar{n}_g}$ and $\bar{\mbf{c}} \in \realset^{n}$ satisfy $X \subseteq \{\bar{\mbf{G}}, \bar{\mbf{c}}\}$, and let $\mbf{m}$ be an interval vector such that $\mbf{m} \supseteq (\mbf{J} - \midpoint{\mbf{J}}) \bar{\mbf{c}}$ and $\midpoint{\mbf{m}} = \bm{0}$. Finally, let $\mbf{P} \in \realsetmat{n}{n}$ be a diagonal matrix defined by $P_{ii} = \rad{m_i} + \sum_{j=1}^{\bar{n}_g} \sum_{k=1}^{m} \rad{J_{ik}} |\bar{G}_{kj}|$ 	
	for all $i=1,2,\dots,n$. Then, $S$ is contained in the \emph{CZ-inclusion} S $\subseteq \gzinclusion(\mbf{J},X) \triangleq \midpoint{\mbf{J}}X \oplus \mbf{P}B_\infty^n.$
\end{theorem}

\begin{remark} \rm \label{rem:czinclusion}
The zonotope $\{\bar{\mbf{G}},\bar{\mbf{c}}\} \supseteq X$ in Theorem \ref{thm:ndyn_czinclusion} is obtained by eliminating all constraints from $X$ according to the algorithm in \cite{Scott2016}, while the interval vector $\mbf{m}$ is obtained using interval arithmetic.
\end{remark}

\section{Joint state and parameter estimation}
\label{sec:jointestimation}

\subsection{Linear systems} \label{sec:jointlinear}

Consider a linear discrete-time system with unknown-but-bounded disturbances and model parameters, given by
\begin{subequations} \label{eq:joint_systemlinear}
	\begin{align}
	\mbf{x}_k & = \mbf{A}\mbf{x}_{k-1} + \mbf{B}_u \mbf{u}_{k-1} + \mbf{B}_p \mbf{p} + \mbf{B}_w \mbf{w}_{k-1}, \label{eq:joint_systemf} \\
	\mbf{y}_k & = \mbf{C} \mbf{x}_{k} + \mbf{D}_u \mbf{u}_{k} + \mbf{D}_p \mbf{p} + \mbf{D}_v \mbf{v}_{k}\label{eq:joint_systemg}.
	\end{align}
\end{subequations}
where $\mbf{x}_k \in \realset^{n}$ is the system state, $\mbf{u}_{k} \in \realset^{n_u}$ is the known input, $\mbf{w}_k \in \realset^{n_w}$ is the process disturbance, $\mbf{y}_k \in \realset^{n_y}$ is the measured output, $\mbf{v}_k \in \realset^{n_v}$ is the measurement disturbance, and $\mbf{p} \in \realset^{n_p}$ are the unknown parameters. In addition, $\mbf{A} \in \realsetmat{n}{n}$, $\mbf{B}_u \in \realsetmat{n}{n_u}$, $\mbf{B}_p \in \realsetmat{n}{n_p}$, $\mbf{B}_w \in \realsetmat{n}{n_w}$, $\mbf{C} \in \realsetmat{n_y}{n}$, $\mbf{D}_u \in \realsetmat{n_y}{n_u}$, $\mbf{D}_d \in \realsetmat{n_y}{n_p}$, and $\mbf{D}_v \in \realsetmat{n_y}{n_v}$. The initial state, model parameters, and disturbances are assumed to be unknown-but-bounded, i.e., $(\mbf{x}_0,\mbf{p},\mbf{w}_{k},\mbf{v}_k) \in X_0 \times P \times  W \times V$, $\forall k \geq 0$, where $X_0$, $P$, $W$, and $V$ are polytopes representable as CZs.

For any $k\geq 0$, the objective is to approximate the solution set of $(\mbf{x}_k,\mbf{p})$ satisfying \eqref{eq:joint_systemlinear}, as accurately as possible, by a guaranteed enclosure $\hat{Z}_k \subset \realset^{n+n_p}$ satisfying $(\mbf{x}_k,\mbf{p}) \in \hat{Z}_k$. We accomplish this here by extending the prediction-update structure proposed in \cite{Scott2016} to a joint state and parameter estimation framework, considering the refinement of the parametric uncertainty $\mbf{p} \in P$. The proposed generalized scheme is given by the following recursion:
\begin{align}
\bar{Z}_k & \supseteq \{ (\mbf{A} \mbf{x}_{k-1} + \mbf{B}_u \mbf{u}_{k-1} + \mbf{B}_p \mbf{p} + \mbf{B}_w \mbf{w}_{k-1}, \mbf{p}) : \nonumber \\
& \quad \quad \quad \quad (\mbf{x}_{k-1}, \mbf{p}) \in \hat{Z}_{k-1}, \,\mbf{w}_{k-1} \in W \}, \label{eq:joint_prediction0}\\
\hat{Z}_k & \supseteq \{ (\mbf{x}_{k},\mbf{p}) \in \bar{Z}_k : \nonumber \\ &  \mbf{C} \mbf{x}_{k} + \mbf{D}_u \mbf{u}_{k} + \mbf{D}_p \mbf{p} + \mbf{D}_v \mbf{v}_{k} = \mbf{y}_k , \, \mbf{v}_{k} \in V \}, \label{eq:joint_update0}
\end{align}
where \eqref{eq:joint_prediction0} is the \emph{joint prediction step}, \eqref{eq:joint_update0} is the \emph{joint update step}, and the scheme is initialized with $\bar{Z}_0 \triangleq \bar{X}_0 \times P$ in the joint update step.
If $\hat{Z}_{k-1}$ is a valid enclosure of $(\mbf{x}_{k-1},\mbf{p})$ for some $k \geq 1$, then $(\mbf{x}_k,\mbf{p}) \in \bar{Z}_k$ given by \eqref{eq:joint_prediction0}. By construction, this leads to $(\mbf{x}_k,\mbf{p}) \in \hat{Z}_k$ from \eqref{eq:joint_update0}.

Exact enclosures for the joint prediction step \eqref{eq:joint_prediction0} and joint update step \eqref{eq:joint_update0} can be obtained straightforwardly using CZs. Since, by assumption, the unknown parameters $\mbf{p}$ are constant, i.e., $\mbf{p}_k = \mbf{p}_{k-1}$, then the prediction and update steps can be computed in CG-rep by defining $\mbf{z}_k \triangleq (\mbf{x}_k, \mbf{p})$, and extending the structure proposed in \cite{Scott2016} to the unified formulation
\begin{align} 
	\bar{Z}_k & = \begin{bmatrix} \mbf{A} & \mbf{B}_p \\ \mbf{0} & \eyenoarg \end{bmatrix} \hat{Z}_{k-1} \oplus \begin{bmatrix} \mbf{B}_u \\ \mbf{0} \end{bmatrix} \mbf{u}_{k-1} \oplus \begin{bmatrix} \mbf{B}_w \\ \mbf{0} \end{bmatrix} W, \label{eq:joint_linearprediction} \\
	\hat{Z}_k & = \bar{Z}_k \cap_{[\mbf{C} \,\; \mbf{D}_p]} ((\mbf{y}_k - \mbf{D}_u \mbf{u}_k) \oplus (-\mbf{D}_v V)). \label{eq:joint_linearupdate}
\end{align}
Note that the enclosure of the parameters $\mbf{p}$ is refined over time through the proposed joint update step \eqref{eq:joint_linearupdate}. In addition, as in linear state estimation using CZs, all the operations in \eqref{eq:joint_linearprediction}--\eqref{eq:joint_linearupdate} can be performed easily using \eqref{eq:pre_czlimage}--\eqref{eq:pre_czintersection}, with linear complexity increase in the number of generators and constraints. Besides, the coupling between states $\mbf{x}_k$ and parameters $\mbf{p}$ is preserved from each time step to the other using the proposed framework. Also, the enclosures in \eqref{eq:joint_linearprediction}--\eqref{eq:joint_linearupdate} are exact if the complexity of the set is not limited. In practice, due to finite computational resources, complexity reduction methods \citep{Scott2016} are used to enclose the sets $\bar{Z}_k$ and $\hat{Z}_k$ by CZs with a desired (lower) number of generators and constraints\footnote{The coupling between states and parameters is maintained even in this case, thus providing benefits with respect to intervals.}.

\subsection{Nonlinear systems with linear output equation} \label{sec:jointnonlinear}

Consider a class of nonlinear discrete-time systems with bounded uncertainties, evolving according to the dynamics
\begin{equation} \label{eq:joint_systemfnonlinear}
    \mbf{x}_k = \mbf{f}(\mbf{x}_{k-1}, \mbf{u}_{k-1}, \mbf{p}, \mbf{w}_{k-1}), 
\end{equation}
and with linear output equation \eqref{eq:joint_systemg}, where the nonlinear function $\mbf{f}: \realset^n \times \realset^{n_u} \times \realset^{n_p} \times \realset^{n_w} \to \realset^n$ is assumed to be of class $\mathcal{C}^1$. The initial state, model parameters, and disturbances are assumed to be unknown-but-bounded, i.e., $(\mbf{x}_0,\mbf{p},\mbf{w}_{k},\mbf{v}_k) \in X_0 \times P \times  W \times V$, $\forall k \geq 0$, where $X_0$, $P$, $W$, and $V$ are polytopes representable as CZs. 
As in the linear case \eqref{eq:joint_systemlinear}, the objective is to enclose the trajectories of \eqref{eq:joint_systemfnonlinear} as accurately as possible by a set $\hat{Z}_k \subset \realset^{n+n_p}$  for any $k\geq 0$. This is accomplished  by extending the method proposed in \cite{Rego2021} to a joint state and parameter estimation framework to allow the refinement of the parameter enclosure $P$ over time, as well as to preserve the couplings between states and parameters. The proposed prediction-update scheme is given by
\begin{equation} \label{eq:joint_prediction0nonlinear}
\begin{aligned}
\bar{Z}_k & \supseteq \{ (\mbf{f}(\mbf{x}_{k-1}, \mbf{u}_{k-1}, \mbf{p}, \mbf{w}_{k-1}), \mbf{p}) 
 \\& \quad \quad \quad \quad
 : (\mbf{x}_{k-1}, \mbf{p}) \in \hat{Z}_{k-1}, \,\mbf{w}_{k-1} \in W \},
\end{aligned}
\end{equation}
as the joint prediction step, and by \eqref{eq:joint_update0} as the joint update step (which remains linear). As in the previous case, the scheme is initialized with $\bar{Z}_0 \triangleq \bar{X}_0 \times P$ in the joint update step.

We first extend the prediction method described by Proposition 1 in \cite{Rego2021} to consider both states and model parameters in a more general framework. This result will be necessary for the nonlinear joint state and parameter estimation method developed in this section. 

The following proposition is based on the Mean Value Theorem, and provides an enclosure for the state $\mbf{x}_k$ in the prediction step \eqref{eq:joint_prediction0nonlinear}. A method to compute an enclosure $\bar{Z}_k$ for the augmented variable $(\mbf{x}_k,\mbf{p})$ satisfying the joint prediction step is given by Corollary \ref{thm:joint_mvejointprediction}, which is derived from the result of Proposition \ref{thm:joint_mvestatepred}.

\begin{proposition} \rm (State prediction) \label{thm:joint_mvestatepred}
	Let $\mbf{f} : \realset^n \times \realset^{n_u} \times \realset^{n_p} \times \realset^{n_w} \to \realset^n$ be of class $\mathcal{C}^1$. Let $\mbf{u} \in \realset^{n_u}$, and let $X\subset \realset^n$, $P \subset \realset^{n_p}$, and $W \subset \realset^{n_w}$ be CZs. Let $Z = X \times P$, and choose any $\bm{\gamma}_z = (\bm{\gamma}_x, \bm{\gamma}_p) \in \square X \times \square P$. If $Z_w$ is a CZ such that $\mbf{f}(\bm{\gamma}_x,\mbf{u},\bm{\gamma}_p,W) \subseteq Z_w$ and $\mbf{J} \in \intvalsetmat{n}{n}$ is an interval matrix satisfying $\nabla^T_z \mbf{f}(\square X,\mbf{u}, \square P,W)\subseteq \mbf{J}$, then $\mbf{f}(X,\mbf{u},P,W) \subseteq Z_w \oplus \gzinclusion\left(\mbf{J},  Z - \bm{\gamma}_z \right)$.
\end{proposition}

\begin{proof} Choose any $(\mathbf{x},\mbf{p},\mathbf{w})\in X\times P \times W$. Let $\mbf{r} \triangleq (\mbf{x},\mbf{p})$, $\bm{\gamma}_r \triangleq (\bm{\gamma}_x,\bm{\gamma}_p)$, $R \triangleq X \times P$. Lemma 1 in \cite{Rego2021}  (with $\bm{\alpha} \triangleq \mbf{f}$, $\mbf{x} \triangleq \mbf{r}$, $\bm{\gamma}_x \triangleq \bm{\gamma}_r$, and $X \triangleq R$) ensures that there exists a real matrix $\hat{\mathbf{J}}\in \mathbf{J}$ such that $\mbf{f}(\mbf{x},\mbf{u},\mbf{p},\mbf{w}) = \mbf{f}(\bm{\gamma}_x,\mbf{u},\bm{\gamma}_p,\mbf{w}) + \hat{\mathbf{J}} (\mathbf{z} - \bm{\gamma}_z)$, with $\mbf{z} = (\mbf{x},\mbf{p})$. By Theorem \ref{thm:ndyn_czinclusion} and the choice of $Z_w$, it follows that $\mbf{f}(\mbf{x},\mbf{u},\mbf{p},\mbf{w}) \in Z_w \oplus \gzinclusion\left(\mathbf{J}, Z - \bm{\gamma}_z\right)$, as desired. \end{proof}

\begin{remark} \rm
    In this note, the interval matrix $\mbf{J}$ is computed by evaluating the analytical expression of $\nabla^T_z \mbf{f}(\square X,\mbf{u}, \square P,W)$ using interval arithmetic. Jacobians can also be computed using factorized formulations of polynomial equations into a quasi-linear form and slope arithmetic. Algorithmic differentiation-based solutions may be found in \cite{Moore2009}.
\end{remark}

\begin{corollary} \rm (Joint prediction) \label{thm:joint_mvejointprediction}
	Let $\mbf{f} : \realset^n \times \realset^{n_u} \times \realset^{n_p} \times \realset^{n_w} \to \realset^n$ be of class $\mathcal{C}^1$. For $k \geq 1$, let: (i) $\mbf{u}_{k-1} \in \realset^{n_u}$, (ii) $\mbf{w}_{k-1} \in W = \{\mbf{G}_w, \mbf{c}_w, \mbf{A}_w, \mbf{b}_w\}$, (iii) $\mbf{p} \in P$, and (iv) $\mbf{z}_{k-1} = (\mbf{x}_{k-1},\mbf{p}) \in \hat{Z}_{k-1} = \{\hat{\mbf{G}}_{k-1}, \hat{\mbf{c}}_{k-1}$, $\hat{\mbf{A}}_{k-1}, \hat{\mbf{b}}_{k-1}\}$. Choose any $\bm{\gamma}_z = (\bm{\gamma}_x, \bm{\gamma}_p) \in \square \hat{Z}_{k-1}$. For all $(\mbf{x}_{k-1},\mbf{p}) \in \square \hat{Z}_{k-1}$, $\mbf{w}_{k-1} \in W$, let: (i) $Z_w$ be a CZ such that $\mbf{f}(\bm{\gamma}_x,\mbf{u}_{k-1},\bm{\gamma}_p,\mbf{w}_{k-1}) \subseteq Z_w$, and (ii) $\mbf{J}_z \in \intvalsetmat{n}{n}$ satisfy $\nabla^T_z \mbf{f}(\mbf{x}_{k-1},\mbf{u}_{k-1}, \mbf{p}, \mbf{w}_{k-1} ) \subseteq \mbf{J}_z$. If $\{\underline{\hat{\mbf{G}}},\underline{\hat{\mbf{c}}}\}$ is a zonotope with $\underline{n}_g$ generators satisfying $\hat{Z}_{k-1} - \bm{\gamma}_z \subseteq \{\underline{\hat{\mbf{G}}},\underline{\hat{\mbf{c}}}\}$, then $(\mbf{x}_k,\mbf{p}) \in \bar{Z}_k$, with
	\begin{equation} \label{eq:joint_mvejointprediction}
    \bar{Z}_k = \begin{bmatrix} \mbf{H} \\ \mbf{E} \end{bmatrix} \hat{Z}_{k-1} \oplus \begin{bmatrix} \mbf{H} \\ \mbf{0} \end{bmatrix} (-\bm{\gamma}_z) \oplus \begin{bmatrix} \hat{\mbf{P}} \\ \mbf{0} \end{bmatrix} B_\infty^n \oplus \begin{bmatrix} \mbf{I} \\ \mbf{0} \end{bmatrix} Z_w, 	
	\end{equation}	
	where $\mbf{E} \triangleq [\zeros{n_p}{n} \,\; \eye{n_p}]$, $\mbf{H} \triangleq \text{mid}(\mbf{J}_z)$, $\hat{\mbf{P}} \in \realsetmat{n}{n}$ is diagonal with $\hat{P}_{ii} = \text{rad}(m_i) + \sum_{j=1}^{\underline{n}_g} \sum_{\ell=1}^{n+n_p} \text{rad}(J_{z,i\ell})|\underline{\hat{G}}_{\ell j}|$, and $\mbf{m} \triangleq (\mbf{J}_z - \text{mid}(\mbf{J}_z)) \underline{\hat{\mbf{c}}} \in \intvalset^n$.
\end{corollary} 
\begin{proof}
Choose any $(\mbf{x}_{k-1},\mbf{p}) = \mbf{z}_{k-1} \in \hat{Z}_{k-1}$, $\mbf{w}_{k-1} \in W$. From \eqref{eq:joint_systemfnonlinear}, Proposition \ref{thm:joint_mvestatepred} and Theorem \ref{thm:ndyn_czinclusion}, there must exist $\bm{\delta} \in B_\infty^n$ such that $\mbf{x}_k = \mbf{f}(\mbf{x}_{k-1},\mbf{u}_{k-1},\mbf{p},\mbf{w}_{k-1}) = \mbf{f}(\bm{\gamma}_x,\mbf{u}_{k-1},\bm{\gamma}_p,\mbf{w}_{k-1}) + \text{mid}(\mbf{J}_z)(\mbf{z}_{k-1} - \bm{\gamma}_z) + \hat{\mbf{P}} \bm{\delta}$, with $\hat{\mbf{P}}$ defined as in the statement of the corollary. Then, by the definition of $\mbf{z}_{k-1}$ and $\mbf{E}$, $\mbf{p} = \mbf{E}\mbf{z}_{k-1}$ holds, and we have that (considering $\mbf{H} \triangleq \text{mid}(\mbf{J}_z)$)
\begin{align*}
(\mbf{x}_{k},\mbf{p}) & = (\mbf{f}(\bm{\gamma}_x,\mbf{u}_{k-1},\bm{\gamma}_p,\mbf{w}_{k-1}) + \mbf{H}(\mbf{z}_{k-1} - \bm{\gamma}_z) + \hat{\mbf{P}} \bm{\delta}, \mbf{E} \mbf{z}_{k-1}) \\
& = (\mbf{H}\mbf{z}_{k-1} - \mbf{H}\bm{\gamma}_z + \hat{\mbf{P}} \bm{\delta} + \mbf{f}(\bm{\gamma}_x,\mbf{u}_{k-1},\bm{\gamma}_p,\mbf{w}_{k-1}), \mbf{E} \mbf{z}_{k-1}) \\
& = \begin{bmatrix} \mbf{H} \\ \mbf{E} \end{bmatrix} \mbf{z}_{k-1} + \begin{bmatrix} \mbf{H} \\ \mbf{0} \end{bmatrix} (-\bm{\gamma}_z) 
\! + \! \begin{bmatrix} \hat{\mbf{P}} \\ \mbf{0} \end{bmatrix} \bm{\delta} \!+ \!\begin{bmatrix} \mbf{I} \\ \mbf{0} \end{bmatrix} \mbf{f}(\bm{\gamma}_x,\mbf{u}_{k-1},\bm{\gamma}_p,\mbf{w}_{k-1}) \\
& \in \begin{bmatrix} \mbf{H} \\ \mbf{E} \end{bmatrix} \hat{Z}_{k-1} \oplus \begin{bmatrix} \mbf{H} \\ \mbf{0} \end{bmatrix} (-\bm{\gamma}_z) \oplus \begin{bmatrix} \hat{\mbf{P}} \\ \mbf{0} \end{bmatrix} B_\infty^{n} \oplus \begin{bmatrix} \mbf{I} \\ \mbf{0} \end{bmatrix} Z_w ,
\end{align*}
with $Z_w$ defined as in the statement of the corollary (this CZ can be obtained analogously to Remark 4 in \cite{Rego2021}), which proves the corollary. \end{proof}

An enclosure of the joint prediction step for the dynamics \eqref{eq:joint_systemfnonlinear} can be obtained in CG-rep using Corollary \ref{thm:joint_mvejointprediction}. As in the linear case, and differently from interval methods, the CZ $\bar{Z}_k$ given by \eqref{eq:joint_mvejointprediction} preserves the existing couplings between state $\mbf{x}_{k-1}$ and parameter $\mbf{p}$. Moreover, due to linearity of the output equation \eqref{eq:joint_systemg}, an exact bound for the update step can be obtained using \eqref{eq:joint_linearupdate}, which in addition refines the enclosure of the parameters $\mbf{p}$, with $\bar{Z}_k$ given by \eqref{eq:joint_mvejointprediction}. 
Bounded enclosures can be obtained only if the condition of full detectability/identifiability of states and parameters is verified. See \cite{paradowski2020observability} for observability analysis in the presence of uncertainty.

\begin{remark} \rm \label{rem:joint_setcomplexity}
	Let the CZs $(\hat{Z}_{k-1}, \bar{Z}_k, W, V)$ have $(\hat{n}_g,\bar{n}_g,n_{g_w},n_{g_v})$ generators, and $(\hat{n}_c,\bar{n}_c,$ $n_{c_w},n_{c_v})$ constraints, respectively. Then, the enclosure obtained by Corollary \ref{thm:joint_mvejointprediction} has $\hat{n}_g + 2n + n_{g_w}$ generators and $\hat{n}_c + n_{c_w}$ constraints. On the other side, the enclosure $\hat{Z}_k$ obtained by \eqref{eq:joint_linearupdate} has $\bar{n}_g + n_{g_v}$ generators, and $\bar{n}_c + n_{c_v} + n_y$ constraints. The computational complexities of all the operations used in this section can be found in \cite{Rego2020}, while the complexities of the proposed method can be derived straightforwardly by replacing $n$ with $n+n_p$ in the expressions obtained in \cite{Rego2020} and \cite{Rego2021}.
\end{remark}

\section{Numerical examples}
	
This section presents numerical results\footnote{The simulations were performed using MATLAB, CPLEX and INTLAB.} for the set-based joint state and parameter estimation method proposed in this note. We compare the results provided by the new framework (denoted by CZ-J for the linear case, and CZMV-J for the nonlinear case) with the CZ methods proposed in \cite{Scott2016} and \cite{Rego2021}, denoted by CZ and CZMV, respectively (i.e., with prediction step given by Proposition \ref{thm:joint_mvestatepred}, in which MV stands for ``Mean Value''), with the interval arithmetic method proposed in \cite{Jaulin2001b}, based on forward-backward propagation (FBP), and the zonotope method proposed in \cite{Alamo2005a}, with intersection operator by Property 1 in \cite{Bravo2006} (Z-J and ZMV-J). Intervals and zonotopes are also applied to the proposed joint estimation framework.

\begin{figure}[!tb]
	\centering{
		\def\svgwidth{0.8\columnwidth}
{\scriptsize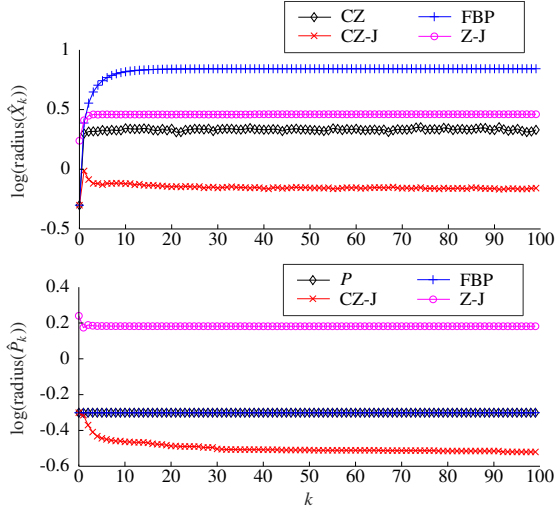}		
    \caption{Linear example for $k \in [0,200]$, the average radii of the projections $\hat{X}_k$ (top) and $\hat{P}_k$ (bottom) of the sets provided by CZ ($\diamond$) and CZ-J ($\times$), Z-J ($\circ$), and FBP ($+$), and the average radii of the original parameter set $P$ ($\diamond$).}\label{fig:linear_state_high_radii}}
\end{figure}

To demonstrate the advantages of performing joint state and parameter estimation using CZ-J, we first consider 10 discrete-time linear systems defined as in \eqref{eq:joint_systemlinear}, with $n=n_w=10$, $n_p=n_v=n_y=6$. The matrices $\mbf{A}$ and $\mbf{C}$ are generated according to a uniform random distribution and satisfy $|a(i,j)|\leq 1/7$ and $|c(i,j)|\leq 1/4 \, , \forall (i,j)$. The matrices $\mbf{B}_p$ and $\mbf{D}_p$ are constituted by values taken from a uniform distribution within $[ -1, 1 ]$. Additionally, $\mbf{B}_w=\eye{n_w}$, $\mbf{D}_v=\eye{n_v}$, $\mbf{B}_u = \zeros{n}{n_u}$ and  $\mbf{D}_u = \zeros{n_y}{n_u}$. 
Process and measurement disturbances satisfy $\|\mbf{w}_k\|_\infty \leq 0.05$ and $\|\mbf{v}_k\|_\infty \leq 0.05$, respectively. The sets $\bar{X_0}$ and $P$ are boxes whose centers are integers randomly selected from $[-6,6]$, according to a uniform discrete distribution, while their radii are $0.5$.
Measurement data have been collected simulating each system and generating process and measurement disturbances according to a uniform random distribution, as well as random initial states, using the listed bounds. The maximum number of generators and constraints of $\hat{Z}_k$ are set to $70$ and $20$, respectively. 
Figure \ref{fig:linear_state_high_radii} shows the average of the radii of the projections $\hat{X}_k$ (top) and $\hat{P}_k$ (bottom) of the enclosures provided by CZ, CZ-J, FBP, and Z-J. 
In this work, only one iteration of FBP was applied, since multiple iterations did not provide a further refinement of the resulting interval enclosures. As it is noticed, the capability to capture the dependence between states and parameters allows CZs to be tighter than intervals. 
CZ-J results in smaller sets w.r.t. CZ and Z-J because the former does not refine the parameter set over time, and the latter relies on conservative intersections with strips. 
In addition to Z-J, in this example, FBP has not even guaranteed a refinement of the parameter enclosure, which is a result from neglecting the dependencies between states and parameters due to the wrapping effect.

Lastly, we present a nonlinear numerical example to demonstrate the effectiveness of the method proposed in Section \ref{sec:jointnonlinear}. Consider the nonlinear discrete-time system described by 
\begin{equation} \label{eq:joint_examplenonlinear}
\begin{aligned}
x_{1,k} & = 3 x_{1,k-1} {-} p x_{1,k-1}^2 {-} \frac{4 x_{1,k-1} x_{2,k-1}}{4 + x_{1,k-1}} + w_{1,k-1} \\
x_{2,k} & = -2 x_{2,k-1} + \frac{3 x_{1,k-1} x_{2,k-1}}{4 + x_{1,k-1}} + w_{2,k-1} \\
y_{1,k} & =  x_{1,k} + v_{1,k}, \quad y_{2,k} = -x_{1,k} + x_{2,k} + (7p-1) + v_{2,k},
\end{aligned}
\end{equation}
with $\|\mbf{w}_k\|_\infty \leq 0.2$, $\|\mbf{v}_k\|_\infty \leq 0.1$, and $p \in P \subset \realset$ being an unknown model parameter. 
The enclosures $\bar{X}_0$ and $P$ are boxes given by $\bar{X}_0 = \{ \text{diag}(1.2, 0.6), (10, 0.5) \}$ and $P = \{ 5, 1/7\}$. 
Measurement data have been obtained by simulating \eqref{eq:joint_examplenonlinear} with $\mbf{x}_0 = (10.2,0.65) \in \bar{X}_0$, and $p = 1/7 \in P$. The process and measurement disturbances are generated from uniform random distributions with the listed bounds. The numbers of generators and constraints of $\hat{Z}_k$ were limited to 8 and 3, respectively. 
\begin{figure}[!tb]
	\centering{
		\def\svgwidth{0.96\columnwidth}
{\scriptsize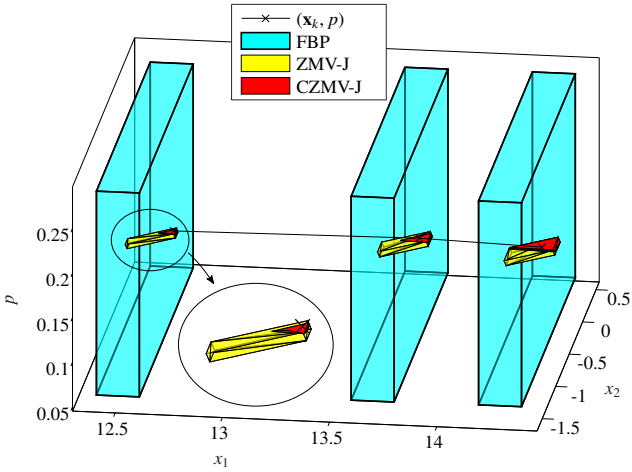}
\caption{The variables $(\mbf{x}_k,p)$ ($\times$), and the sets obtained using ZMV-J (yellow), CZMV-J (red), and interval arithmetic (cyan), for the nonlinear system \eqref{eq:joint_examplenonlinear} at $k\in\{3,15,149\}$.}\label{fig:linear13D}} 
\end{figure}
\begin{figure}[!tb]
	\centering{
		\def\svgwidth{0.9\columnwidth}
{\scriptsize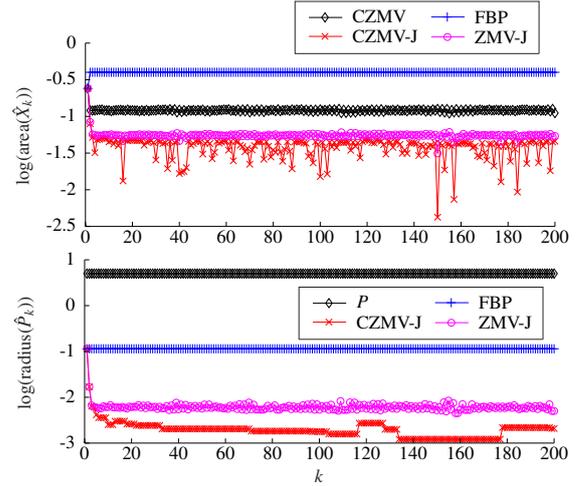}		
        \caption{Areas of the projections $\hat{X}_k$ (top) and radii of $\hat{P}_k$ (bottom) provided by CZMV ($\diamond$) and CZMV-J ($\times$), FBP ($+$), ZMV-J ($\circ$), and the radius of the original parameter set $P$ ($\diamond$) for the nonlinear system \eqref{eq:joint_examplenonlinear}.}\label{fig:nonlinearexvolume}} 
\end{figure}

Figure \ref{fig:linear13D} shows the sets $\hat{Z}_k$ obtained using FBP, ZMV-J, and CZMV-J, as well as the evolution of $\mbf{x}_k$, for $k\in\{3,15,149\}$. As in the linear case, the state enclosures provided by CZMV-J are significantly smaller than the ones obtained with FBP, due to capability to capture the dependencies between states and parameters. Figure \ref{fig:nonlinearexvolume}, which shows the areas of $\hat{X}_k$ (top) and the radii of $\hat{P}_k$ (bottom), as well as the results obtained using ZMV-J (the CZMV-J to ZMV-J average area ratio of $\hat{X}_k$ is $71\%$, while the average radius ratio of $\hat{P}_k$ is $34\%$), corroborates the advantages of using the proposed joint framework CZMV-J. 

\section{Conclusions}
\label{sec:conclusions}

This note developed a new method for set-based joint state and parameter estimation of discrete-time systems with unknown-but-bounded model parameters. By extending state estimation methods using CZs to a unified framework, allowing to maintain the dependencies between states and parameters, the accuracy of both state and parameter estimation was significantly improved. Future works will include extending the method developed in Section \ref{sec:jointlinear} to joint state and parameter estimation of linear descriptor systems.









\bibliography{\bibfolder/masterthesis_bib,\bibfolder/appendices_bib,\bibfolder/UAVControl_bib,\bibfolder/BackgroundHist_bib,\bibfolder/Surveys_bib,\bibfolder/PassiveFTC_bib,\bibfolder/ActiveFTC_bib,\bibfolder/UAVFTC,\bibfolder/SetTheoretic_bib,\bibfolder/SetTheoreticFTCFDI_bib,\bibfolder/Davide_bib,\bibfolder/paperAutomatica_bib,\bibfolder/paperCDC_bib,\bibfolder/paperECC_bib,\bibfolder/paperIFAC_bib,\bibfolder/paperNonlinearMeas_bib,\bibfolder/Robotic_bib,\bibfolder/stelios_bibliography,\bibfolder/phdthesis_bib,\bibfolder/paperParameter_bib,\bibfolder/Diego_bib}

\begin{thebibliography}{16}
\expandafter\ifx\csname natexlab\endcsname\relax\def\natexlab#1{#1}\fi
\providecommand{\url}[1]{\texttt{#1}}
\providecommand{\href}[2]{#2}
\providecommand{\path}[1]{#1}
\providecommand{\DOIprefix}{doi:}
\providecommand{\ArXivprefix}{arXiv:}
\providecommand{\URLprefix}{URL: }
\providecommand{\Pubmedprefix}{pmid:}
\providecommand{\doi}[1]{\href{http://dx.doi.org/#1}{\path{#1}}}
\providecommand{\Pubmed}[1]{\href{pmid:#1}{\path{#1}}}
\providecommand{\bibinfo}[2]{#2}
\ifx\xfnm\relax \def\xfnm[#1]{\unskip,\space#1}\fi
\bibitem[{Alamo et~al.(2005)Alamo, Bravo \& Camacho}]{Alamo2005a}
\bibinfo{author}{Alamo, T.}, \bibinfo{author}{Bravo, J.}, \&
  \bibinfo{author}{Camacho, E.} (\bibinfo{year}{2005}).
\newblock \bibinfo{title}{Guaranteed state estimation by zonotopes}.
\newblock {\it \bibinfo{journal}{Automatica}\/},  {\it \bibinfo{volume}{41}\/},
  \bibinfo{pages}{1035--1043}.
\bibitem[{Bravo et~al.(2006)Bravo, Alamo \& Camacho}]{Bravo2006}
\bibinfo{author}{Bravo, J.~M.}, \bibinfo{author}{Alamo, T.}, \&
  \bibinfo{author}{Camacho, E.~F.} (\bibinfo{year}{2006}).
\newblock \bibinfo{title}{Bounded error identification of systems with
  time-varying parameters}.
\newblock {\it \bibinfo{journal}{IEEE Transactions on Automatic Control}\/},
  {\it \bibinfo{volume}{51}\/}, \bibinfo{pages}{1144--1150}.
\bibitem[{Chisci et~al.(1996)Chisci, Garulli \& Zappa}]{Chisci1996}
\bibinfo{author}{Chisci, L.}, \bibinfo{author}{Garulli, A.}, \&
  \bibinfo{author}{Zappa, G.} (\bibinfo{year}{1996}).
\newblock \bibinfo{title}{{Recursive state bounding by parallelotopes}}.
\newblock {\it \bibinfo{journal}{Automatica}\/},  {\it \bibinfo{volume}{32}\/},
  \bibinfo{pages}{1049--1055}.
\bibitem[{Cui et~al.(2020)Cui, Ding, Jin, Alsaedi \& Hayat}]{Cui2020}
\bibinfo{author}{Cui, T.}, \bibinfo{author}{Ding, F.}, \bibinfo{author}{Jin,
  X.-B.}, \bibinfo{author}{Alsaedi, A.}, \& \bibinfo{author}{Hayat, T.}
  (\bibinfo{year}{2020}).
\newblock \bibinfo{title}{Joint multi-innovation recursive extended least
  squares parameter and state estimation for a class of state-space systems}.
\newblock {\it \bibinfo{journal}{International Journal of Control, Automation
  and Systems}\/},  {\it \bibinfo{volume}{18}\/}, \bibinfo{pages}{1412--1424}.
\bibitem[{Denis-Vidal et~al.(2019)Denis-Vidal, Jauberthie \&
  Kieffer}]{Vidal2019}
\bibinfo{author}{Denis-Vidal, L.}, \bibinfo{author}{Jauberthie, C.}, \&
  \bibinfo{author}{Kieffer, M.} (\bibinfo{year}{2019}).
\newblock \bibinfo{title}{Optimal experiment design for bounded-error
  estimation of nonlinear models}.
\newblock In {\it \bibinfo{booktitle}{Proc. of the 58th Conference on Decision
  and Control}\/} (pp. \bibinfo{pages}{4147--4154}).
\bibitem[{Jaulin et~al.(2001)Jaulin, Braems, Kieffer \& Walter}]{Jaulin2001b}
\bibinfo{author}{Jaulin, L.}, \bibinfo{author}{Braems, I.},
  \bibinfo{author}{Kieffer, M.}, \& \bibinfo{author}{Walter, E.}
  (\bibinfo{year}{2001}).
\newblock \bibinfo{title}{Nonlinear state estimation using forward-backward
  propagation of intervals in an algorithm}.
\newblock In {\it \bibinfo{booktitle}{Scientific Computing, Validated Numerics,
  Interval Methods}\/} (pp. \bibinfo{pages}{191--201}).
\newblock \bibinfo{publisher}{Springer}.
\bibitem[{K{\"u}hn(1998)}]{Kuhn1998}
\bibinfo{author}{K{\"u}hn, W.} (\bibinfo{year}{1998}).
\newblock \bibinfo{title}{Rigorously computed orbits of dynamical systems
  without the wrapping effect}.
\newblock {\it \bibinfo{journal}{Computing}\/},  {\it \bibinfo{volume}{61}\/},
  \bibinfo{pages}{47--67}.
\bibitem[{Moore et~al.(2009)Moore, Kearfott \& Cloud}]{Moore2009}
\bibinfo{author}{Moore, R.~E.}, \bibinfo{author}{Kearfott, R.~B.}, \&
  \bibinfo{author}{Cloud, M.~J.} (\bibinfo{year}{2009}).
\newblock {\it \bibinfo{title}{Introduction to Interval Analysis}\/}.
\newblock \bibinfo{address}{Philadelphia, PA, USA}: \bibinfo{publisher}{SIAM}.
\bibitem[{Paradowski et~al.(2020)Paradowski, Lerch, Damaszek, Dehnert \&
  Tibken}]{paradowski2020observability}
\bibinfo{author}{Paradowski, T.}, \bibinfo{author}{Lerch, S.},
  \bibinfo{author}{Damaszek, M.}, \bibinfo{author}{Dehnert, R.}, \&
  \bibinfo{author}{Tibken, B.} (\bibinfo{year}{2020}).
\newblock \bibinfo{title}{Observability of uncertain nonlinear systems using
  interval analysis}.
\newblock {\it \bibinfo{journal}{Algorithms}\/},  {\it \bibinfo{volume}{13}\/},
  \bibinfo{pages}{66}.
\bibitem[{Ra{\i}ssi et~al.(2004)Ra{\i}ssi, Ramdani \& Candau}]{raissi2004set}
\bibinfo{author}{Ra{\i}ssi, T.}, \bibinfo{author}{Ramdani, N.}, \&
  \bibinfo{author}{Candau, Y.} (\bibinfo{year}{2004}).
\newblock \bibinfo{title}{Set membership state and parameter estimation for
  systems described by nonlinear differential equations}.
\newblock {\it \bibinfo{journal}{Automatica}\/},  {\it \bibinfo{volume}{40}\/},
  \bibinfo{pages}{1771--1777}.
\bibitem[{Rego et~al.(2020)Rego, Raffo, Scott \& Raimondo}]{Rego2020}
\bibinfo{author}{Rego, B.~S.}, \bibinfo{author}{Raffo, G.~V.},
  \bibinfo{author}{Scott, J.~K.}, \& \bibinfo{author}{Raimondo, D.~M.}
  (\bibinfo{year}{2020}).
\newblock \bibinfo{title}{Guaranteed methods based on constrained zonotopes for
  set-valued state estimation of nonlinear discrete-time systems}.
\newblock {\it \bibinfo{journal}{Automatica}\/},  {\it
  \bibinfo{volume}{111}\/}, \bibinfo{pages}{108614}.
\bibitem[{Rego et~al.(2021)Rego, Scott, Raimondo \& Raffo}]{Rego2021}
\bibinfo{author}{Rego, B.~S.}, \bibinfo{author}{Scott, J.~K.},
  \bibinfo{author}{Raimondo, D.~M.}, \& \bibinfo{author}{Raffo, G.~V.}
  (\bibinfo{year}{2021}).
\newblock \bibinfo{title}{Set-valued state estimation of nonlinear
  discrete-time systems with nonlinear invariants based on constrained
  zonotopes}.
\newblock {\it \bibinfo{journal}{Automatica}\/},  {\it
  \bibinfo{volume}{129}\/}, \bibinfo{pages}{109628}.
\bibitem[{Rumschinski et~al.(2010)Rumschinski, Borchers, Bosio, Weismantel \&
  Findeisen}]{Rumschinski2010}
\bibinfo{author}{Rumschinski, P.}, \bibinfo{author}{Borchers, S.},
  \bibinfo{author}{Bosio, S.}, \bibinfo{author}{Weismantel, R.}, \&
  \bibinfo{author}{Findeisen, R.} (\bibinfo{year}{2010}).
\newblock \bibinfo{title}{Set-based dynamical parameter estimation and model
  invalidation for biochemical reaction networks}.
\newblock {\it \bibinfo{journal}{BMC Systems Biology}\/},  {\it
  \bibinfo{volume}{4}\/}, \bibinfo{pages}{1--14}.
\bibitem[{Scott et~al.(2016)Scott, Raimondo, Marseglia \& Braatz}]{Scott2016}
\bibinfo{author}{Scott, J.~K.}, \bibinfo{author}{Raimondo, D.~M.},
  \bibinfo{author}{Marseglia, G.~R.}, \& \bibinfo{author}{Braatz, R.~D.}
  (\bibinfo{year}{2016}).
\newblock \bibinfo{title}{Constrained zonotopes: a new tool for set-based
  estimation and fault detection}.
\newblock {\it \bibinfo{journal}{Automatica}\/},  {\it \bibinfo{volume}{69}\/},
  \bibinfo{pages}{126--136}.
\bibitem[{Wang et~al.(2017)Wang, Kolmanovsky \& Sun}]{Wang2017}
\bibinfo{author}{Wang, H.}, \bibinfo{author}{Kolmanovsky, I.}, \&
  \bibinfo{author}{Sun, J.} (\bibinfo{year}{2017}).
\newblock \bibinfo{title}{Zonotope-based set-membership parameter
  identification of linear systems with additive and multiplicative
  uncertainties: A new algorithm}.
\newblock In {\it \bibinfo{booktitle}{Proc. of the 2017 ACC}\/} (pp.
  \bibinfo{pages}{1481--1486}).
\bibitem[{Zhang et~al.(2020)Zhang, Couto, Park, Gill \& Moura}]{Zhang2020}
\bibinfo{author}{Zhang, D.}, \bibinfo{author}{Couto, L.~D.},
  \bibinfo{author}{Park, S.}, \bibinfo{author}{Gill, P.}, \&
  \bibinfo{author}{Moura, S.~J.} (\bibinfo{year}{2020}).
\newblock \bibinfo{title}{Nonlinear state and parameter estimation for li-ion
  batteries with thermal coupling}.
\newblock In {\it \bibinfo{booktitle}{Proc. of the 21st IFAC World Congress}\/}
  (pp. \bibinfo{pages}{12658--12663}).

\end{thebibliography}

\end{document}